\documentclass[conference]{IEEEtran}

\usepackage{cite}
\usepackage{amsmath,amssymb,amsfonts}
\usepackage{algorithmic}
\usepackage{algorithm}
\usepackage{multirow}
\usepackage{graphicx}
\usepackage{textcomp}
\usepackage{xcolor, soul, enumitem}
\usepackage{booktabs}
\usepackage{microtype}

\def\BibTeX{{\rm B\kern-.05em{\sc i\kern-.025em b}\kern-.08em
    T\kern-.1667em\lower.7ex\hbox{E}\kern-.125emX}}

\begin{document}


\title{Watermarking Neuromorphic Brains: Intellectual Property Protection in Spiking Neural Networks}

\author{\IEEEauthorblockN{Hamed Poursiami}
\IEEEauthorblockA{\textit{ECE Department} \\ \textit{George Mason University} \\
Fairfax, VA, USA\\
Email: hpoursia@gmu.edu}
\and
\IEEEauthorblockN{Ihsen Alouani}
\IEEEauthorblockA{\textit{Centre for Secure Information Technologies} \\ \textit{Queen's University Belfast} \\
Belfast, UK\\
Email: i.alouani@qub.ac.uk}
\and
\IEEEauthorblockN{ Maryam Parsa}
\IEEEauthorblockA{\textit{ECE Department} \\ \textit{George Mason University} \\
Fairfax, VA, USA\\
Email: mparsa@gmu.edu}
}

\maketitle
\begin{abstract}
As spiking neural networks (SNNs) gain traction in deploying neuromorphic computing solutions, protecting their intellectual property (IP) has become crucial. Without adequate safeguards, proprietary SNN architectures are at risk of theft, replication, or misuse, which could lead to significant financial losses for the owners. While IP protection techniques have been extensively explored for artificial neural networks (ANNs), their applicability and effectiveness for the unique characteristics of SNNs remain largely unexplored. In this work, we pioneer an investigation into adapting two prominent watermarking approaches, namely, fingerprint-based and backdoor-based mechanisms to secure proprietary SNN architectures. We conduct thorough experiments to evaluate the impact on fidelity, resilience against overwrite threats, and resistance to compression attacks when applying these watermarking techniques to SNNs, drawing comparisons with their ANN counterparts. This study lays the groundwork for developing neuromorphic-aware IP protection strategies tailored to the distinctive dynamics of SNNs.
\end{abstract}

\section{Introduction}
\label{sec:intro}

\noindent
Spiking neural networks (SNNs) have emerged as a promising biologically-inspired computing paradigm that offers potential advantages over traditional artificial neural networks (ANNs) in terms of energy efficiency and ability to process spatio-temporal data \cite{roy2019towards, christensen20222022}. As SNNs gain traction in research and commercial applications, there is a growing need to protect the intellectual property (IP) of these intricate models.
The development of {state-of-the-art SNN architectures} \cite{zhu2023spikegpt,zhang2022recent} is often a computationally-intensive endeavor that requires massive datasets \cite{amir2017low,mitrokhin2019ev}, considerable domain expertise, and immense computing resources over extended time periods. Without adequate IP protection measures in place, these proprietary SNN models run the risk of being stolen, replicated, or misused by adversaries, leading to potential financial losses and legal challenges for the model owners. Safeguarding SNN IP has therefore become a critical concern as these models transition from academic research into real-world deployments across diverse fields such as computer vision, speech recognition, and robotics \cite{kim2021optimizing, wu2020deep,bing2018survey}.

While techniques for IP protection of ANNs have been extensively studied \cite{peng2023intellectual}, the applicability of these methods and their effectiveness to the emerging field of neuromorphic computing remains largely unexplored. To the best of our knowledge, this is the {first work} exploring IP protection for SNNs. We investigate the feasibility of adopting model watermarking approaches from ANNs to SNNs, and evaluating their performance against potential IP threats in the SNN domain.

IP protection scenarios can be divided into white-box and black-box settings based on the verifier's perspective \cite{regazzoni2021protecting}. In a white-box scenario, the verifier has access to the internal parameters of the model, while in a black-box scenario, verification is performed only through interactions with an unauthorized service provider via the prediction API.
Moreover, the protection mechanisms can be broadly categorized based on their implementation as parameter-based, backdoor-based, and fingerprinting-based \cite{xue2021intellectual}. Parameter-based methods involve embedding a unique marker directly into the model's weights or parameters \cite{uchida2017embedding}. Backdoor-based techniques use hidden triggers within the model to encode the watermark information, inspired by methods used in backdoor attacks \cite{zhang2018protecting,szyller2021dawn}. Fingerprinting-based approaches utilize existing features of the model to create a unique fingerprint for IP protection\cite{lukas2019deep}.
Given that pirated models are often deployed as online services accessible only through remote APIs, we focus our attention on the black-box scenario. Consequently, as embedding parameter-based watermarks requires white-box access to the model weights, we concentrate our investigation on exploring backdoor-based and fingerprint-based mechanisms.

\noindent The key contributions of this work are as follows:
\begin{enumerate}[leftmargin=5mm]

    \item We propose an investigation of IP protection for SNNs. Specifically, we adapt two prominent watermarking approaches to suit the unique characteristics of SNNs.

    \item Through extensive experiments, we assess the fidelity impact of watermarking SNNs and their resilience against adaptive threats models.

    \item We provide insights into the effectiveness of existing ANN-dedicated watermarking schemes when applied to SNNs, paving the way for the development of tailored IP protection techniques for neuromorphic computing models.
\end{enumerate}
\section{Preliminary: Spiking Neural Networks}
\label{sec:preliminary}

\noindent

Unlike traditional ANNs that use continuous activation values, SNNs communicate through discrete spikes over time, similar to how real neurons fire\cite{schuman2022opportunities}. In an SNN, each neuron accumulates incoming spikes from other connected neurons. When this accumulated signal crosses a threshold, the neuron generates an output spike that gets passed along to other neurons. The leaky integrate-and-fire (LIF) model is one of the most widely used approaches for simulating this spiking behavior. 

It models the phenomenon where a neuron integrates incoming spikes over time, while its membrane potential gradually decreases due to leakage of charge. The neuron resets its membrane potential after firing a spike\cite{izhikevich2004model}.
The dynamics of a single LIF neuron can be mathematically expressed in discrete-time as follows:
\begin{equation}
    \nu [n] = \alpha \cdot \nu [n-1] + \sum_{k}^{}\omega_k \cdot I _{\text{k}}[n] - O[n-1] \cdot \eta
\label{eq:membrance}
\end{equation}

Where $\nu$ represents the membrane potential of the spiking neuron, $n$ denotes the discretized time steps, and $\alpha$ is the leakage factor that models the gradual decay of the potential over time. The term $I _{\text{k}}$ corresponds to the input spikes arriving from the presynaptic neuron $k$, weighted by the synaptic strength $\omega_k$. The activation function $O[n]$ is defined as follows:

\begin{equation}
O[n]=\left\{\begin{matrix}
1 & ,if \ \nu[n]>\eta\\ 
0 & ,otherwise.
\end{matrix}\right.
\label{eq:activation}
\end{equation}

This neuron model adopts a soft-reset mechanism, which involves subtracting the threshold $\eta$ from the membrane potential upon spike generation. To process static data with SNNs, inputs must be encoded into spike trains using schemes like rate encoding, where higher values correspond to higher firing rates \cite{gerstner2014neuronal}.
Training SNNs can leverage methods similar to backpropagation through time (BPTT) for sequential models\cite{bellec2018long}. However, BPTT faces challenges due to the non-differentiable spiking activation function. A solution is to substitute this function with a smooth, differentiable surrogate during the backward pass of the training process \cite{neftci2019surrogate}.

\section{Watermarking SNNs}
\label{sec:methodology}

\noindent
In this section, we explore the applicability of two prominent intellectual property (IP) protection techniques, originally designed for ANNs, to the emerging field of SNNs. Specifically, we investigate the fingerprinting-based approach by \cite{le2020adversarial} and the backdoor-based technique by \cite{adi2018turning}. 
\\ Our objective is to assess the feasibility and effectiveness of adapting these techniques to the unique characteristics of SNNs, such as their spiking nature, temporal dynamics, and training methodologies.  In the following subsections, we first provide a brief overview of each technique and then discuss our approach to implementing them in the context of SNNs.

\subsection{Fingerprint-based}

\noindent One of the most well-known fingerprinting-based approaches known as ``adversarial frontier snitching" is proposed by Merrer et al.\cite{le2020adversarial}, which utilizes adversarial examples as watermark keys. Adversarial examples, or adversaries in short, are inputs crafted by applying small perturbations to mislead a machine learning model. 
A widely used method to generate adversaries is the Fast Gradient Sign Method (FGSM), which computes the adversary \( X' \) as:
\begin{equation}
X' = X + \epsilon \cdot sign(\nabla_X L(X, y))
\label{eq:fgsm}
\end{equation}

\noindent Where \( X \) is the original input, \( y \) is the true label, \( L(X, y) \) denotes the loss function used for model training, and \( \epsilon \) controls the perturbation strength.

\begin{figure}[tb!]
    \centering
    \includegraphics[width=\linewidth]{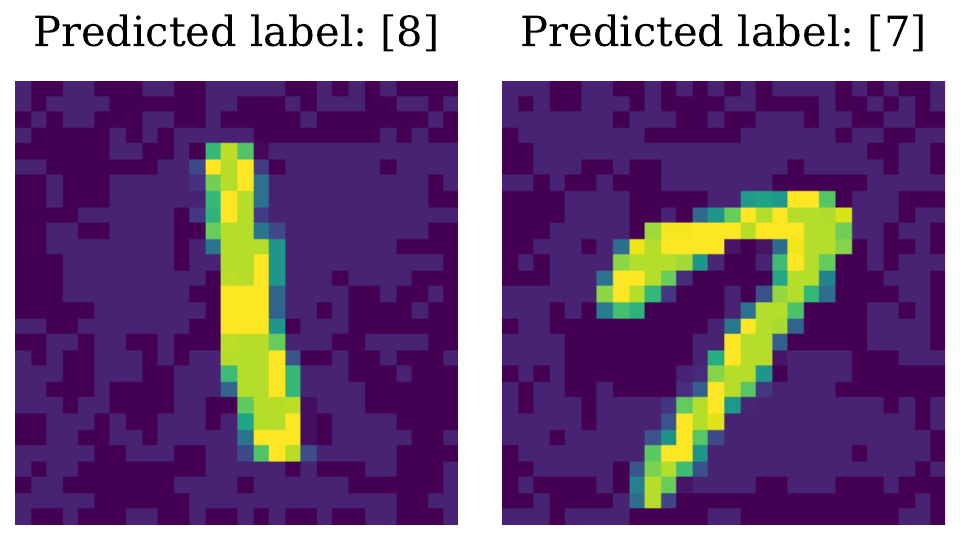}
    \caption{Examples of adversaries generated by FGSM. Left: True adversaries, Right: False adversaries} 

    \label{fig:adversary}
\end{figure}

\begin{figure*}[tb!]
    \centering
    \includegraphics[width=\linewidth]{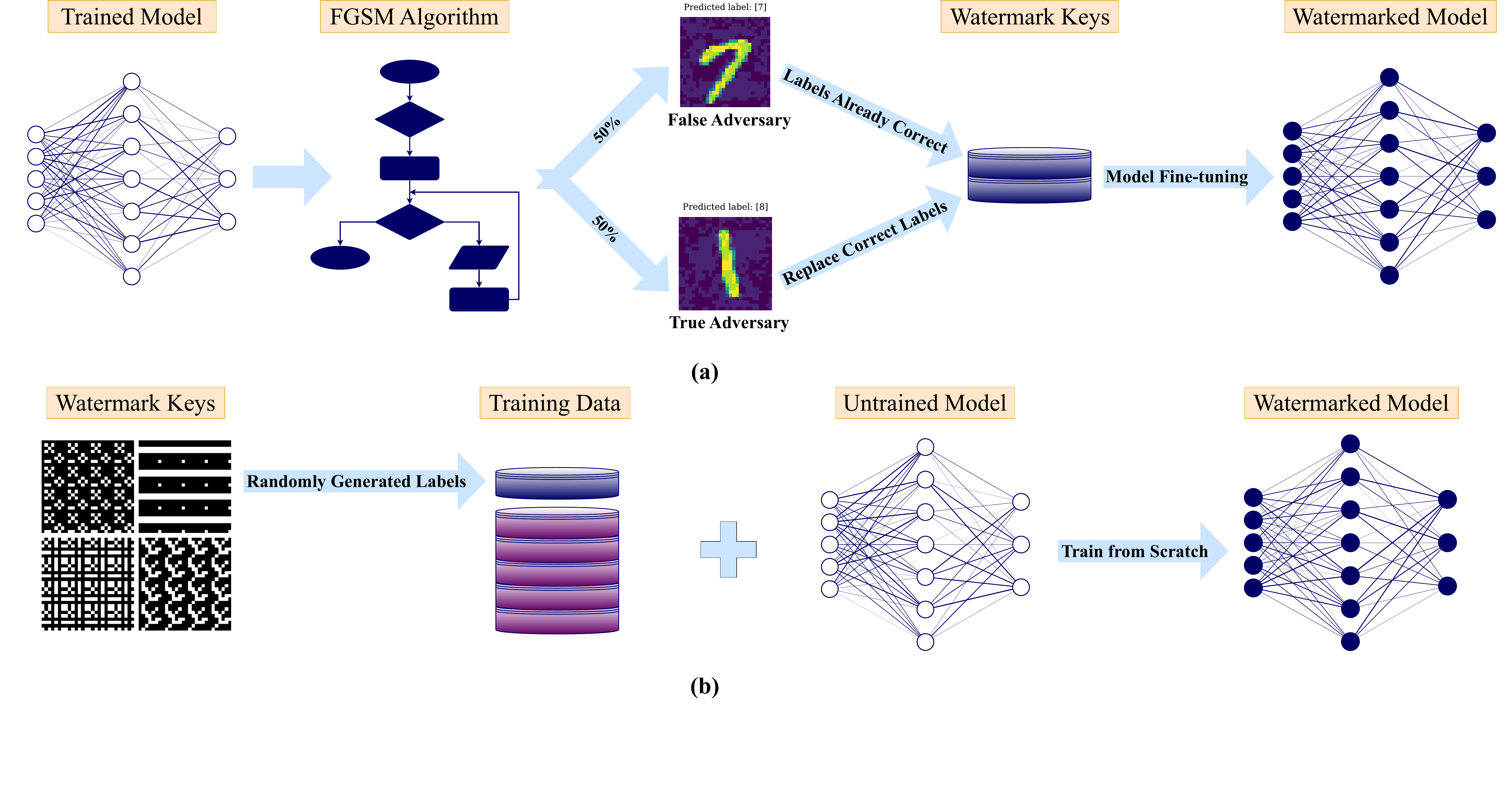}
    \caption{Workflows of the (a) fingerprint-based adversarial frontier stitching approach \cite{le2020adversarial} and (b) backdoor-based watermarking technique \cite{adi2018turning}} 

    \label{fig:process}
\end{figure*}

The adversarial frontier stitching algorithm involves generating two types of adversarial examples: ``true adversaries" that are misclassified by the original model, and ``false adversaries" that are correctly classified but close to the decision boundary. Figure \ref{fig:adversary} illustrates examples of true and false adversarial examples. The model is then fine-tuned so that the true adversaries become correctly classified while the false ones remain close to the decision boundary. During this fine-tuning process, the true adversaries adjust the decision boundaries of the model to make it unique, essentially embedding a fingerprint. The false adversaries limit the extent to which the decision boundaries get perturbed when incorporating the true adversary labels. Figure \ref{fig:process}a depicts the overall procedure of this method.
\\ To detect this watermark in a potentially pirated model, one can query it with the set of watermark key images (true and false adversaries) and check if the predictions match the expected labels. If the number of mismatches is below a threshold, it indicates the presence of the watermark.

Adapting the adversarial frontier stitching technique to SNNs presents unique obstacles. While the gradients required for generating adversaries (as per equation 3) are typically computed using backpropagation, the non-differentiable nature of the LIF neuron activations (equation 2) hinders this process. To overcome this, we employ a smooth, differentiable surrogate function during the backward pass to enable gradient flow. In our experiments, we utilize the fast Sigmoid function \cite{zenke2021brain} as the surrogate for both training and adversary generation via FGSM.
However, a more fundamental challenge arises from the inherent characteristics of spiking inputs. Directly applying continuous perturbations to the binary spike trains, as suggested by equation 3, would violate their binary encoding and result in invalid representations. 
It is worth noting that we also cannot backpropagate gradients from the spiking inputs to the static input domain. This is because the rate encoding scheme used to convert static inputs into spike trains is a stochastic and non-reversible procedure \cite{poursiami2024brainleaks}. Therefore, propagating gradients in the reverse direction is infeasible.

To mitigate these challenges, we approximate the static input $X$ by the time average of the spiking input $I_{in}$, denoted as  $\bar{I}_{in}$, computed as:
\begin{equation}
\bar{I}_{in} = \frac{1}{T}\sum_{n=1}^{T}I_{in}[n]
\end{equation}
\\ Here, $T$ represents the total number of time steps.
Subsequently, for the sign of the gradients, we have:

\begin{equation}
sign(\nabla_X L) \approx sign(\nabla_{\bar{I}_{in}} L)
\end{equation}
\\ For convolutional SNNs, authors in \cite{sharmin2020inherent} have derived a relationship to calculate $sign(\nabla_{\bar{I}_{\text{in}}} L)$ based on the gradients of the loss with respect to the membrane potentials of the first layer $\nu_{(1)}$, expressed as:

\begin{equation} 
    sign(\nabla_{\bar{I}_{in}} L) = sign\left (\nabla_{\nu_{(1)}} L \ \star \  W_{(1)}^{flipped}\right)
\end{equation}
\\ Here, $W_{(1)}^{flipped}$ denotes the flipped kernel (weight matrix rotated 180 degrees) corresponding to the first layer, and $\star$ is the convolution operation. 
\\ Following the same steps, we can utilize their approach to derive $sign(\nabla_{\bar{I}_{\text{in}}} L)$ for fully-connected SNNs as:
\begin{equation}
    sign(\nabla_{\bar{I}_{in}} L) = sign\left( W_{(1)}^{T} \ \cdot \  \nabla_{\nu_{(1)}} L\right)
\end{equation}
\\ Where, $W_{(1)}^{T}$ represents the transpose of the first-layer weight matrix. Algorithm \ref{alg:1} provides the details of the adversary generation procedure via the FGSM for SNNs and  CSNNs.

\subsection{Backdoor-based}

\noindent Adi et al. \cite{adi2018turning} proposed a backdoor-based approach for watermarking deep neural networks along with leveraging a cryptographic commitment scheme to enable tracking the intellectual property of pre-trained models. The key idea is to embed watermarks into the DNN models during training, enabling the models to learn and memorize the patterns of the embedded watermarks. The authors use a set of abstract images as the watermarks along with randomly assigned labels and optimize the model on both the original dataset and the trigger set either by training from scratch or by fine-tuning the model. This process is illustrated in Figure \ref{fig:process}b. The watermarked model can then be verified remotely by sending queries containing the watermarks and checking if the models produce the pre-defined predictions associated with the watermarks.

Applying this backdoor-based watermarking to SNNs is straightforward  for several reasons. Firstly, it does not inherently rely on the differentiability of activations, thus avoiding the challenges posed by the non-differentiable LIF neurons in SNNs. Additionally, by encoding the watermark patterns as spike train representations, this method can operate directly on the binary spike inputs used by SNNs.

\begin{algorithm}[t!]
\caption{SNN FGSM}
\begin{algorithmic}[1]
 \renewcommand{\algorithmicrequire}{\textbf{Input:}}
 \renewcommand{\algorithmicensure}{\textbf{Initialize:}}
 
 \REQUIRE SNN Model ($M$), Total number of time-steps ($T$), Original Sample $(X, y)$, Perturbation Strength ($\epsilon$) 
 \ENSURE $I_{in} = RateEncoder(X)$ \\ \hspace{25pt} $\nabla = 0$

 \FOR {n = 1 to T}

 \STATE $L = Loss \left ( M\left (I_{in}\left [n  \right ]  \right ),y  \right )$

 \STATE $\nabla  \ += \nabla_{\nu_{(1)}} L \left ( M\left (I_{in}\left [n  \right ]  \right ),y  \right )$
 
 \ENDFOR

 \IF{Model is CNN}
 \STATE $X' = X + \epsilon \cdot sign \ (\ \nabla \ \cdot \  W_{(1)}^{flipped}\ )$
 \ELSIF{Model is FC}
 \STATE $X' = X + \epsilon \cdot sign\ ( \ W_{(1)}^{T} \ \cdot \  \nabla \ )$
 \ENDIF
 
 \RETURN $X'$ 
\end{algorithmic} 
\label{alg:1}
\end{algorithm}
\section{Threat Model}
\label{sec:Threat}

\noindent \textbf{Attacker objective:} Intellectual property (IP) protection mechanisms, such as watermarking techniques, must be resilient against various threats that an attacker might employ to overwrite or remove the embedded watermarks. In this work, we investigate two primary categories of threats: overwriting threats and compression threats. These threats represent scenarios where an adversary, with white-box access to a victim model, attempts to distort or erase the watermark patterns embedded within it.

\subsection{Overwriting Attacks}

\noindent Overwriting threats involve further training the pirated model on additional data, aiming to adjust the decision boundaries and potentially overwrite the watermark information. These attacks are considered highly probable as they are commonly employed and demands less computational resources and training data. We explore two variants:

\begin{enumerate}
    \item \textbf{Model Fine-tuning:} The attacker continues training the pirated model using a small dataset sampled from the same or a similar distribution as the original data. 
    
    \item \textbf{Adversarial Fine-tuning:} The attacker crafts adversarial examples to target and overwrite the watermark patterns. By exploiting the model's vulnerabilities, the attacker attempts to force the model to "unlearn" the watermark information during fine-tuning.
\end{enumerate}

\subsection{Compression Attacks}
\noindent
Model compression techniques aim to reduce the computational and memory requirements of neural networks. However, these methods can potentially distort or erase the embedded watermark patterns, posing a threat to IP protection. In this work, we investigate two compression strategies:

\begin{enumerate}
    \item \textbf{L1 Pruning:} Weights with small magnitudes, based on their L1 norms, are removed by setting them to zero. This approach eliminates the contribution of low-magnitude weights to the model's computations.
    
    \item \textbf{Random Pruning:} Weights are randomly selected and pruned (set to zero), regardless of their magnitude.
\end{enumerate}

\section{Experiments}
\label{sec:experiments}


\noindent
\textit{a) Dataset and Models:}  We investigate IP protection schemes on digit classification task using the MNIST dataset \cite{6296535}. The pirated model employs a fully connected network with two hidden layers, each comprising 512 neurons. To gauge false positives, we also utilize an innocent model with a single hidden layer containing 3000 neurons. In our experiments, we sample a subset of 1000 images from the MNIST test set to perform model fine-tuning attack. For the adversarial fine-tuning threat, we generate adversaries corresponding to the same subset using the FGSM method with $\epsilon=0.1$ and $\epsilon=0.2$. The accuracy of the fine-tuned models is then evaluated using the remaining 9000 images from the test set. 

\vspace{6pt}
\noindent
\textit{b) Training Specifications:}  Transitioning to the SNN paradigm, we replace Sigmoid activations with LIF neuron with a leakage rate of 0.7. Training employs BPTT with a fast-Sigmoid function (slope: 40) for surrogate gradient calculation. The loss function incorporates cumulative Softmax cross-entropy applied to the membrane potentials of the output layer across all time-steps.
Data preprocessing involves transforming static images into 25-step spiking representations using a rate-encoding scheme to ensure the compatibility with the neuromorphic architecture.
All experiments in this study were conducted using PyTorch framework, with SNNtorch \cite{eshraghian2023training} enabling incorporation of spiking neuron models.

\vspace{6pt}
\noindent
\textit{c) Evaluation Metrics:} We Assess the robustness of watermarking techniques against the threats outlined in the section \ref{sec:Threat} by considering \textit{{model validation accuracy}} and the \textit{watermark misclassification tally}. Model validation accuracy reflects how well a watermarked or pirated model maintains its predictive performance, indicating the fidelity impact of watermarking or attacks. The watermark misclassification tally is utilized to confirm the presence of the watermark in a potentially pirated model. During verification, the watermarked model is queried with the set of watermark keys (true and false adversaries or backdoor patterns), and the number of mismatches between the model's predictions and the expected labels is computed. A lower watermark misclassification tally below a threshold signifies the presence of the watermark.

To determine this threshold, we adopt the probabilistic approach proposed by \cite{le2020adversarial}, which estimates the likelihood of an unmarked model producing correct predictions on the watermark keys. Specifically, the threshold $\theta$ can be obtained from: 
\begin{equation}
    2^{-\left | K \right |}\sum_{z=0}^{\theta}\binom{\left | K \right |}{z} \leq p
\end{equation}
Where $\left | K \right |$ is the size of the watermark set and $p$ is the desired significance level (p-value). For our experiments with $\left | K \right |=20$ (fingerprint-based) and $\left | K \right |=50$ (backdoor-based), the resulting thresholds for a p-value of $0.05$ are $\theta=6$ and $\theta=19$ mismatches, respectively. If the number of mismatches fall below $\theta$, the model is deemed watermarked.

\begin{table*}[tbh!]
\caption{Comparison of the effectiveness of the  Frontier Snitching watermarking on ANN and SNN models}
\resizebox{\linewidth}{!}{%
\begin{tabular}{@{}ccccccc@{}}
\toprule
\multirow{2}{*}{\textbf{Model}}                      & \multicolumn{3}{c}{\textbf{ANN}} & \multicolumn{3}{c}{\textbf{SNN}} \\
                                                     & Acc     & Mismatched   & Verify  & Acc     & Mismatched   & Verify  \\ \midrule
\textbf{Original}                                    & 98.21   & 10           & False   & 96.75   & 10           & False   \\
\textbf{Watermarked}                                 & 97.37   & 0            & True    & \textbf{90.12}   & 1            & True    \\
\textbf{Innocent}                                    & 96.30   & 7            & False   & 96.39   & 11           & False   \\
\textbf{Adversarial Fine-tuning $[\epsilon = 0.1]$} & 97.30   & 2            & True    & 94.01   & 1            & True    \\
\textbf{Adversarial Fine-tuning $[\epsilon = 0.2]$} & 96.51   & 3            & True    & 93.31   & 8            & \textbf{False}   \\
\textbf{Non-Adversarial Fine-tuning }                 & 98.12   & 3            & True    & 94.26   & 6            & \textbf{False}   \\ \bottomrule
\end{tabular}%
}

\label{tab:merrer}
\end{table*}
\begin{figure}[b!]
    \centering
    \includegraphics[width=\linewidth]{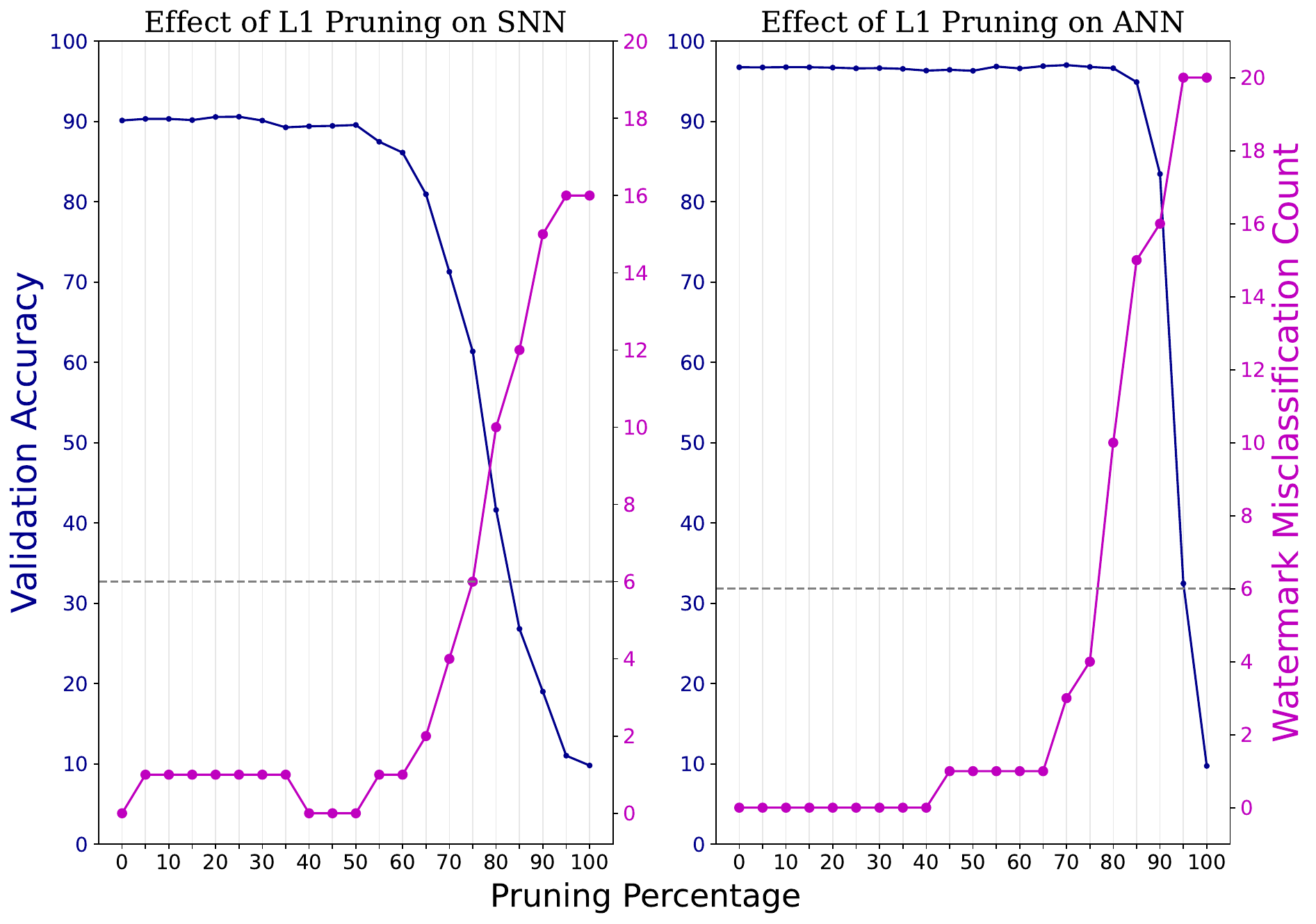}
    \caption{L1 Pruning threat on fingerprint-based watermarked ANN and SNN models. 
    }
    \label{fig:merrer_L1}
\end{figure}

\subsection{Fingerprint-based}

\begin{figure}[b!]
    \centering
    \includegraphics[width=\linewidth]{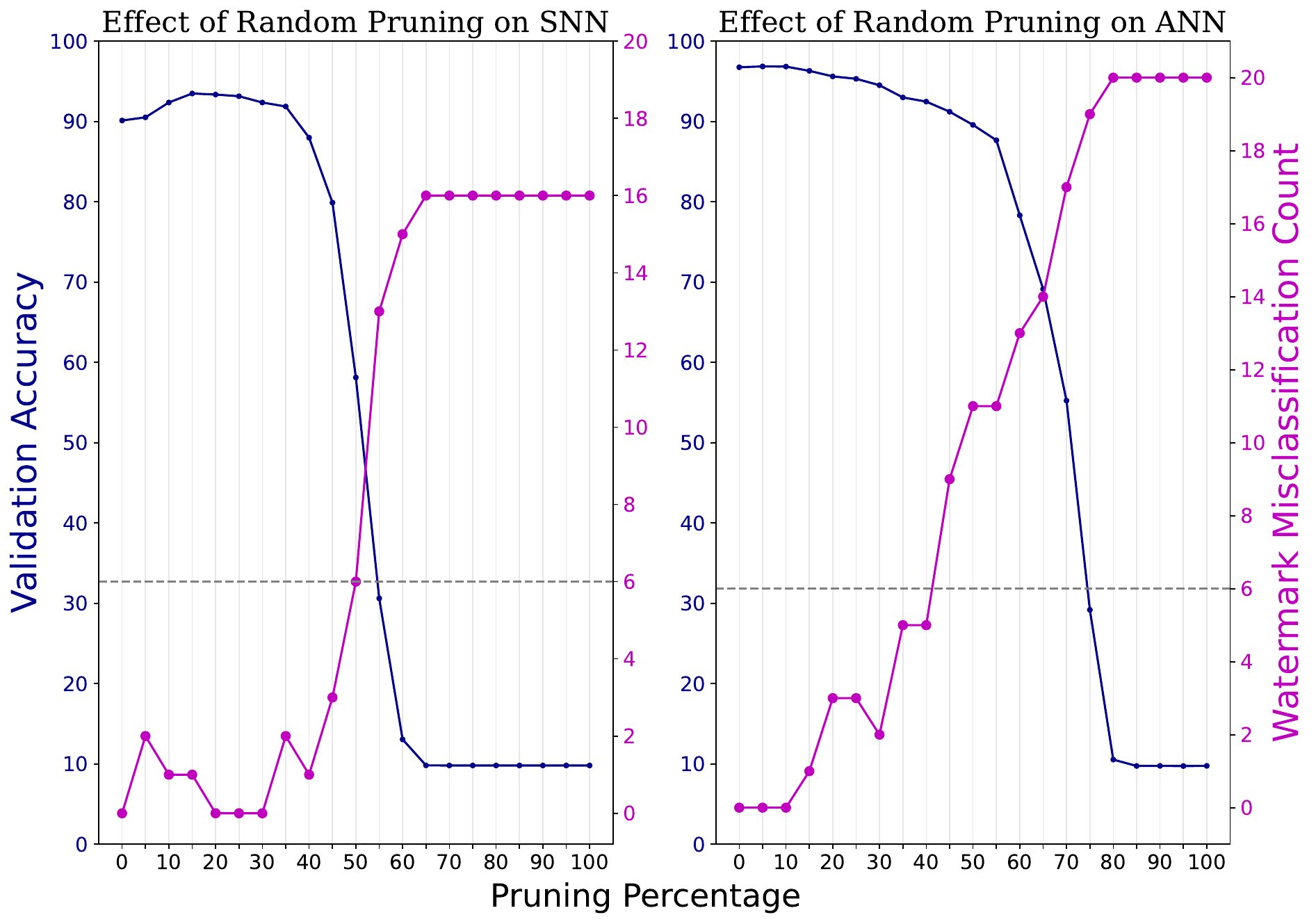}
    \caption{Random Pruning threat on fingerprint-based watermarked ANN and SNN models.%
    }
    \label{fig:merrer_rand}
\end{figure}

\noindent Following the methodology of \cite{le2020adversarial}, we incorporate 10 true adversaries and 10 false adversaries as our model watermark keys. The model is then fine-tuned to memorize (overfit) these watermarks.

The model validation accuracies, along with the results of the overwriting threats, are presented in Table \ref{tab:merrer} for both ANNs and SNNs. The watermarking process has a more significant negative impact on the performance of SNNs compared to ANNs. For ANNs, the accuracy decreases from 98.21\% to 97.37\%, a relatively small drop of 0.84\%. However, SNNs experience a substantial accuracy degradation from 96.75\% to 90.12\%, a reduction of 6.63\%. This observation suggests that embedding watermarks through the adversarial frontier stitching method can remarkably impair the fidelity of SNN models, potentially rendering this technique less effective for watermarking SNNs compared to ANNs.

Furthermore, the results demonstrate SNNs' higher susceptibility to watermark overwriting attacks. While the watermarks are well-preserved during overwriting attacks in ANNs, they are effectively overwritten in SNNs through non-adversarial model fine-tuning and adversarial fine-tuning with $\epsilon=0.2$, highlighting the lower robustness of embedded watermarks in SNNs. However, it is worth noting that the innocent model is correctly identified as unmarked in both ANNs and SNNs, indicating a low likelihood of false positives.

The robustness against compression threats is illustrated in Figures \ref{fig:merrer_L1} and \ref{fig:merrer_rand}, depicting the impact of L1 pruning and random pruning attacks, respectively. It can be observed that watermark elimination comes at a greater performance cost for SNNs. For ANNs, the watermark is removed after 80\% L1 pruning, while maintaining 96.63\% accuracy. In contrast, for SNNs, the watermark is removed after 75\% L1 pruning, with the model accuracy significantly dropping to 61.36\%, rendering the pruned model ineffective. A similar trend is observed for random pruning, where the SNN model degrades to 58.12\% accuracy after watermark removal at 50\% pruning, while the ANN model retains 91.21\% accuracy after 45\% pruning. These results indicate SNNs' increased robustness against pruning threats compared to ANNs, which can be attributed to the inherent redundancy and distributed nature of spiking neural representations.

\begin{table*}[t!]
\caption{Comparison of the effectiveness of the  backdoor-based watermarking on ANN and SNN models}
\resizebox{\linewidth}{!}{%
\begin{tabular}{@{}ccccccc@{}}
\toprule
\multirow{2}{*}{\textbf{Model}}                      & \multicolumn{3}{c}{\textbf{ANN}} & \multicolumn{3}{c}{\textbf{SNN}} \\
                                                     & Acc     & Mismatched   & Verify  & Acc     & Mismatched   & Verify  \\ \midrule
\textbf{Original}                                    & 98.21   & 45           & False   & 96.75   & 45           & False   \\
\textbf{Watermarked}                                 & 98.12   & 0            & True    & \textbf{95.03}   & 0            & True    \\
\textbf{Innocent}                                    & 96.30   & 48           & False   & 96.39   & 44           & False   \\
\textbf{Adversarial Fine-tuning $[\epsilon = 0.1]$} & 96.19   & 16           & True    & 93.87   & 36           & \textbf{False}   \\
\textbf{Adversarial Fine-tuning $[\epsilon = 0.2]$} & 94.42   & 23           & False   & 94.08   & 46           & False   \\
\textbf{Non-Adversarial Fine-tuning}                 & 97.89   & 0            & True    & 94.01   & 0            & True    \\ \bottomrule
\end{tabular}%
}

\label{tab:adi}
\end{table*}

\subsection{Backdoor-based}

\begin{figure}[b!]
    \centering
    \includegraphics[width=\linewidth]{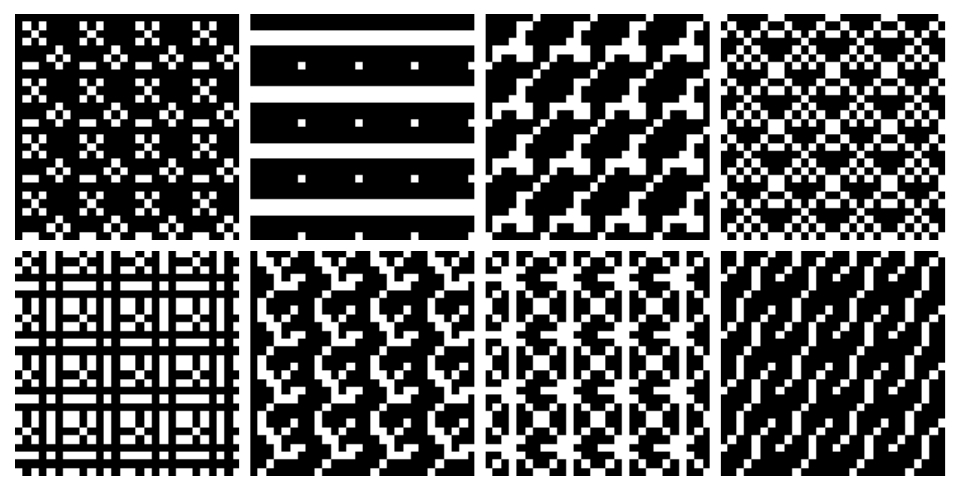}
    \caption{Eight examples of abstract patterns generated as watermarks for backdoor-based approach.} 
    \label{fig:abstract}
\end{figure}

\noindent To evaluate the backdoor-based watermarking technique \cite{adi2018turning}, we generate 50 unique abstract patterns to serve as watermark triggers. These patterns are randomly assigned labels and exhibit no correlation with the MNIST dataset or among themselves. Consequently, knowledge of a subset of patterns does not provide any information to an adversary about the remaining watermark patterns. Examples of these abstract watermark patterns are illustrated in Figure \ref{fig:abstract}.

We incorporate these watermark triggers during the model training process from scratch, as this approach yields better watermarking results compared to fine-tuning a pre-trained model. Table \ref{tab:adi} summarizes the results for both ANN and SNN models. The watermarking process has a negligible impact on the validation accuracy for ANNs, with only a minor drop from 98.21\% to 98.12\%. For SNNs, the accuracy decreases from 96.75\% to 95.03\%, a more substantial reduction compared to ANNs, but still maintaining reasonably high performance. Moreover, the verification results on the innocent model indicate a low likelihood of false positives for this watermarking approach.

When assessing the overwriting threats, we find that the embedded backdoor watermarks exhibit robust persistence against non-adversarial fine-tuning attacks in both ANNs and SNNs. However, the watermarks demonstrate higher vulnerability to adversarial fine-tuning, particularly in SNNs. For a perturbation strength of $\epsilon = 0.1$, the watermark remains intact in ANNs with 16 mismatches out of 50, while it is effectively overwritten in SNNs, as indicated by the high number of mismatches (36 out of 50) during verification. 

\begin{figure}[t!]
    \centering
    \includegraphics[width=\linewidth]{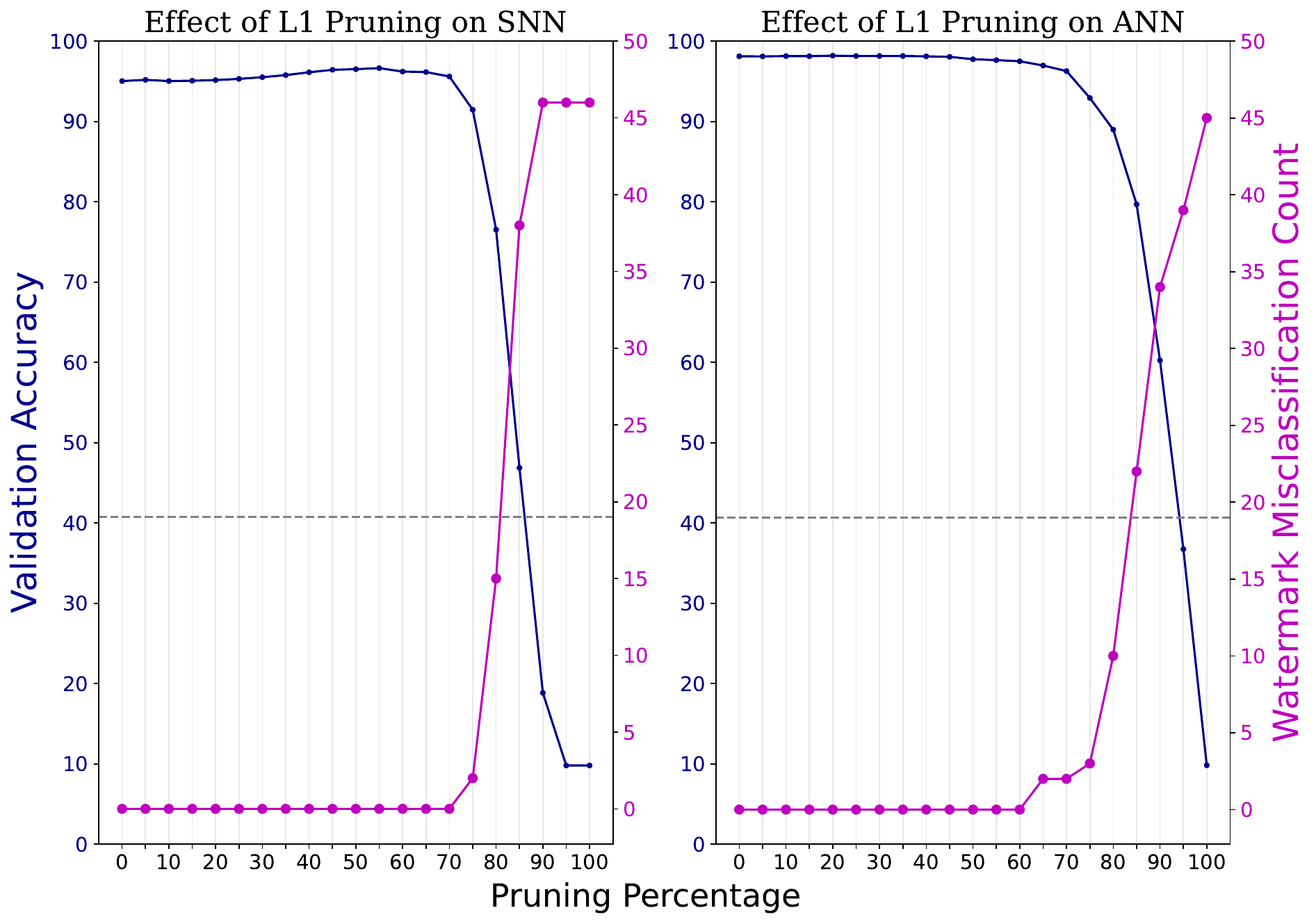}
    \caption{L1 Pruning threat on backdoor-based watermarked ANN and SNN models}
    \label{fig:adi_L1}
\end{figure}

\begin{figure}[t!]
    \centering
    \includegraphics[width=\linewidth]{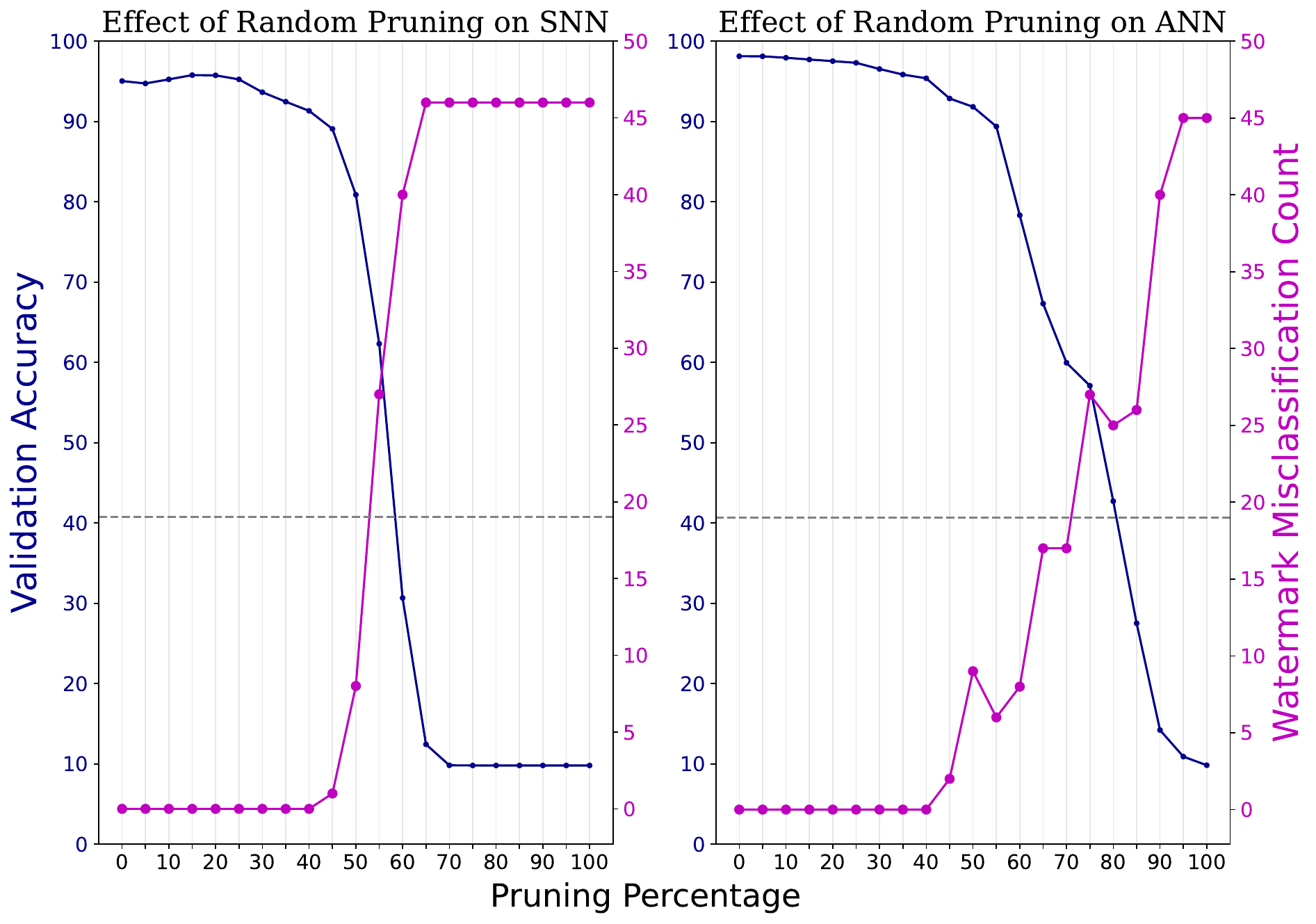}
    \caption{Random Pruning threat on backdoor-based watermarked ANN and SNN models}
    \label{fig:adi_rand}
\end{figure}

Figures \ref{fig:adi_L1} and \ref{fig:adi_rand} assess the resilience against L1 and random pruning attacks. In both ANNs and SNNs, eliminating the watermarks significantly degrades model performance, rendering them ineffective. This underscores the robustness of the backdoor watermarking scheme against compression threats, as the removal of watermarks via pruning substantially impairs model functionality while preserving the embedded intellectual property protection.

  

\subsection{Discussion}

\noindent The experimental results from this study reveal that, in general, SNNs face more performance degradation through the watermarking process compared to ANNs.
However, the performance degradation observed with the fingerprint-based approach is more pronounced, which restricts the applicability of this watermarking scheme for SNNs. In contrast, the backdoor-based approach exhibits reasonable fidelity, with a comparatively minor impact on model accuracy during the watermark embedding process. This difference can be attributed to the nature of backdoor watermarks, which are introduced as patterns outside the original data distribution, potentially altering the decision boundaries in obsolete regions of the high-dimensional space without significantly affecting the model's core functionality on the primary task.

Furthermore, a noticeable observation across both watermarking techniques is the higher susceptibility of SNNs to overwriting attacks compared to ANNs, potentially due to the temporal dynamics and distributed nature of their representations. In SNNs, information is encoded and spread over multiple time steps, introducing additional degrees of freedom during fine-tuning that could make it easier to overwrite or distort the embedded watermark patterns.

Notably, the results indicate that SNNs exhibit greater resilience against compression threats than ANNs. Specifically, the higher dimensionality of SNNs necessitates a larger number of parameters to construct the decision boundaries, rendering them more sensitive to weight pruning. Consequently, as evidenced by the experiments, significant performance degradation in SNNs occurs at lower pruning percentages across all scenarios, owing to their lower degree of overparameterization compared to ANNs.

Overall, these findings suggest that while the backdoor-based watermarking technique is better suited for IP protection in SNNs compared to the fingerprint-based approach, it still exhibits lower fidelity and increased vulnerability against overwriting attacks in SNNs relative to ANNs. Therefore, future research is needed to explore watermarking methods tailored specifically for SNNs, taking into account their unique characteristics and dynamics.

\section{Conclusion}
\label{sec:conclusion}

\noindent
This study explored the applicability of watermarking techniques, originally developed for ANNs, to the emerging domain of SNNs. We investigated two of the most prominent methods within fingerprint-based and backdoor-based mechanisms, by applying necessary modifications to adapt these techniques for SNNs. Our findings reveal that although the backdoor-based approach generally fits better with SNNs,  both techniques have exhibit higher vulnerability to overwriting attacks and a decreased of fidelity in SNNs compared to ANNs. This can be attributed to the unique temporal dynamics and distributed nature of spiking representations. Interestingly, SNNs demonstrated greater resilience against weight pruning threats, owing to their higher dimensionality and lower degree of overparameterization. Overall, this work lays the foundation for future research in exploring neuromorphic-aware watermarking and IP protection strategies tailored to the unique characteristics of SNNs.

\nocite{*}
\bibliographystyle{IEEEtran}
\bibliography{ref}

\begin{thebibliography}{10}
\providecommand{\url}[1]{#1}
\csname url@samestyle\endcsname
\providecommand{\newblock}{\relax}
\providecommand{\bibinfo}[2]{#2}
\providecommand{\BIBentrySTDinterwordspacing}{\spaceskip=0pt\relax}
\providecommand{\BIBentryALTinterwordstretchfactor}{4}
\providecommand{\BIBentryALTinterwordspacing}{\spaceskip=\fontdimen2\font plus
\BIBentryALTinterwordstretchfactor\fontdimen3\font minus \fontdimen4\font\relax}
\providecommand{\BIBforeignlanguage}[2]{{%
\expandafter\ifx\csname l@#1\endcsname\relax
\typeout{** WARNING: IEEEtran.bst: No hyphenation pattern has been}%
\typeout{** loaded for the language `#1'. Using the pattern for}%
\typeout{** the default language instead.}%
\else
\language=\csname l@#1\endcsname
\fi
#2}}
\providecommand{\BIBdecl}{\relax}
\BIBdecl

\bibitem{roy2019towards}
K.~Roy, A.~Jaiswal, and P.~Panda, ``Towards spike-based machine intelligence with neuromorphic computing,'' \emph{Nature}, vol. 575, no. 7784, pp. 607--617, 2019.

\bibitem{christensen20222022}
D.~V. Christensen, R.~Dittmann, B.~Linares-Barranco, A.~Sebastian, M.~Le~Gallo, A.~Redaelli, S.~Slesazeck, T.~Mikolajick, S.~Spiga, S.~Menzel \emph{et~al.}, ``2022 roadmap on neuromorphic computing and engineering,'' \emph{Neuromorphic Computing and Engineering}, vol.~2, no.~2, p. 022501, 2022.

\bibitem{zhu2023spikegpt}
R.-J. Zhu, Q.~Zhao, G.~Li, and J.~K. Eshraghian, ``Spikegpt: Generative pre-trained language model with spiking neural networks,'' \emph{arXiv preprint arXiv:2302.13939}, 2023.

\bibitem{zhang2022recent}
D.~Zhang, S.~Jia, and Q.~Wang, ``Recent advances and new frontiers in spiking neural networks,'' \emph{arXiv preprint arXiv:2204.07050}, 2022.

\bibitem{amir2017low}
A.~Amir, B.~Taba, D.~Berg, T.~Melano, J.~McKinstry, C.~Di~Nolfo, T.~Nayak, A.~Andreopoulos, G.~Garreau, M.~Mendoza \emph{et~al.}, ``A low power, fully event-based gesture recognition system,'' in \emph{Proceedings of the IEEE conference on computer vision and pattern recognition}, 2017, pp. 7243--7252.

\bibitem{mitrokhin2019ev}
A.~Mitrokhin, C.~Ye, C.~Ferm{\"u}ller, Y.~Aloimonos, and T.~Delbruck, ``Ev-imo: Motion segmentation dataset and learning pipeline for event cameras,'' in \emph{2019 IEEE/RSJ International Conference on Intelligent Robots and Systems (IROS)}.\hskip 1em plus 0.5em minus 0.4em\relax IEEE, 2019, pp. 6105--6112.

\bibitem{kim2021optimizing}
Y.~Kim and P.~Panda, ``Optimizing deeper spiking neural networks for dynamic vision sensing,'' \emph{Neural Networks}, vol. 144, pp. 686--698, 2021.

\bibitem{wu2020deep}
J.~Wu, E.~Y{\i}lmaz, M.~Zhang, H.~Li, and K.~C. Tan, ``Deep spiking neural networks for large vocabulary automatic speech recognition,'' \emph{Frontiers in neuroscience}, vol.~14, p. 513257, 2020.

\bibitem{bing2018survey}
Z.~Bing, C.~Meschede, F.~R{\"o}hrbein, K.~Huang, and A.~C. Knoll, ``A survey of robotics control based on learning-inspired spiking neural networks,'' \emph{Frontiers in neurorobotics}, vol.~12, p.~35, 2018.

\bibitem{peng2023intellectual}
S.~Peng, Y.~Chen, J.~Xu, Z.~Chen, C.~Wang, and X.~Jia, ``Intellectual property protection of dnn models,'' \emph{World Wide Web}, vol.~26, no.~4, pp. 1877--1911, 2023.

\bibitem{regazzoni2021protecting}
F.~Regazzoni, P.~Palmieri, F.~Smailbegovic, R.~Cammarota, and I.~Polian, ``Protecting artificial intelligence ips: a survey of watermarking and fingerprinting for machine learning,'' \emph{CAAI Transactions on Intelligence Technology}, vol.~6, no.~2, pp. 180--191, 2021.

\bibitem{xue2021intellectual}
M.~Xue, Y.~Zhang, J.~Wang, and W.~Liu, ``Intellectual property protection for deep learning models: Taxonomy, methods, attacks, and evaluations,'' \emph{IEEE Transactions on Artificial Intelligence}, vol.~3, no.~6, pp. 908--923, 2021.

\bibitem{uchida2017embedding}
Y.~Uchida, Y.~Nagai, S.~Sakazawa, and S.~Satoh, ``Embedding watermarks into deep neural networks,'' in \emph{Proceedings of the 2017 ACM on international conference on multimedia retrieval}, 2017, pp. 269--277.

\bibitem{zhang2018protecting}
J.~Zhang, Z.~Gu, J.~Jang, H.~Wu, M.~P. Stoecklin, H.~Huang, and I.~Molloy, ``Protecting intellectual property of deep neural networks with watermarking,'' in \emph{Proceedings of the 2018 on Asia conference on computer and communications security}, 2018, pp. 159--172.

\bibitem{szyller2021dawn}
S.~Szyller, B.~G. Atli, S.~Marchal, and N.~Asokan, ``Dawn: Dynamic adversarial watermarking of neural networks,'' in \emph{Proceedings of the 29th ACM International Conference on Multimedia}, 2021, pp. 4417--4425.

\bibitem{lukas2019deep}
N.~Lukas, Y.~Zhang, and F.~Kerschbaum, ``Deep neural network fingerprinting by conferrable adversarial examples,'' \emph{arXiv preprint arXiv:1912.00888}, 2019.

\bibitem{schuman2022opportunities}
C.~D. Schuman, S.~R. Kulkarni, M.~Parsa, J.~P. Mitchell, P.~Date, and B.~Kay, ``Opportunities for neuromorphic computing algorithms and applications,'' \emph{Nature Computational Science}, vol.~2, no.~1, pp. 10--19, 2022.

\bibitem{izhikevich2004model}
E.~M. Izhikevich, ``Which model to use for cortical spiking neurons?'' \emph{IEEE transactions on neural networks}, vol.~15, no.~5, pp. 1063--1070, 2004.

\bibitem{gerstner2014neuronal}
W.~Gerstner, W.~M. Kistler, R.~Naud, and L.~Paninski, \emph{Neuronal dynamics: From single neurons to networks and models of cognition}.\hskip 1em plus 0.5em minus 0.4em\relax Cambridge University Press, 2014.

\bibitem{bellec2018long}
G.~Bellec, D.~Salaj, A.~Subramoney, R.~Legenstein, and W.~Maass, ``Long short-term memory and learning-to-learn in networks of spiking neurons,'' \emph{Advances in neural information processing systems}, vol.~31, 2018.

\bibitem{neftci2019surrogate}
E.~O. Neftci, H.~Mostafa, and F.~Zenke, ``Surrogate gradient learning in spiking neural networks: Bringing the power of gradient-based optimization to spiking neural networks,'' \emph{IEEE Signal Processing Magazine}, vol.~36, no.~6, pp. 51--63, 2019.

\bibitem{le2020adversarial}
E.~Le~Merrer, P.~Perez, and G.~Tr{\'e}dan, ``Adversarial frontier stitching for remote neural network watermarking,'' \emph{Neural Computing and Applications}, vol.~32, no.~13, pp. 9233--9244, 2020.

\bibitem{adi2018turning}
Y.~Adi, C.~Baum, M.~Cisse, B.~Pinkas, and J.~Keshet, ``Turning your weakness into a strength: Watermarking deep neural networks by backdooring,'' in \emph{27th USENIX Security Symposium (USENIX Security 18)}, 2018, pp. 1615--1631.

\bibitem{zenke2021brain}
F.~Zenke and E.~O. Neftci, ``Brain-inspired learning on neuromorphic substrates,'' \emph{Proceedings of the IEEE}, vol. 109, no.~5, pp. 935--950, 2021.

\bibitem{poursiami2024brainleaks}
H.~Poursiami, I.~Alouani, and M.~Parsa, ``Brainleaks: On the privacy-preserving properties of neuromorphic architectures against model inversion attacks,'' \emph{arXiv preprint arXiv:2402.00906}, 2024.

\bibitem{sharmin2020inherent}
S.~Sharmin, N.~Rathi, P.~Panda, and K.~Roy, ``Inherent adversarial robustness of deep spiking neural networks: Effects of discrete input encoding and non-linear activations,'' in \emph{Computer Vision--ECCV 2020: 16th European Conference, Glasgow, UK, August 23--28, 2020, Proceedings, Part XXIX 16}.\hskip 1em plus 0.5em minus 0.4em\relax Springer, 2020, pp. 399--414.

\bibitem{6296535}
L.~Deng, ``The mnist database of handwritten digit images for machine learning research [best of the web],'' \emph{IEEE Signal Processing Magazine}, vol.~29, no.~6, pp. 141--142, 2012.

\bibitem{eshraghian2023training}
J.~K. Eshraghian, M.~Ward, E.~O. Neftci, X.~Wang, G.~Lenz, G.~Dwivedi, M.~Bennamoun, D.~S. Jeong, and W.~D. Lu, ``Training spiking neural networks using lessons from deep learning,'' \emph{Proceedings of the IEEE}, 2023.

\end{thebibliography}
\end{document}